\documentclass[aps,preprint]{revtex4}%
\usepackage{amsfonts}
\usepackage{amsmath}
\usepackage{amssymb}
\usepackage{graphicx}%
\providecommand{\U}[1]{\protect\rule{.1in}{.1in}}
\newtheorem{theorem}{Theorem}
\newtheorem{acknowledgement}[theorem]{Acknowledgement}

\begin{document}
\title{Spectroscopic tests for short-range modifications of Newtonian and
post-Newtonian potentials}
\author{A. S. Lemos$^{2,3}$}
\author{G. C. Luna$^{1}$}
\author{E. Maciel$^{3}$}
\author{F. Dahia$^{1}$}
\affiliation{$^{1}$Department of Physics, Federal University of Para\'{\i}ba - Jo\~{a}o
Pessoa -PB - Brazil }
\affiliation{$^{2}$Dep. of Phys., State University of Para\'{\i}ba - Campina Grande - PB - Brazil.}
\affiliation{$^{3}$Department of Physics, Federal University of Campina Grande - Campina
Grande -PB - Brazil}

\pacs{PACS number}

\begin{abstract}
There are theoretical frameworks, such as the large extra dimension models,
which predict the strengthening of the gravitational field in short distances.
Here we obtain new empiric constraints for deviations of standard gravity in
the atomic length scale from analyses of recent and accurate data of hydrogen
spectroscopy. The new bounds, extracted from $1S-3S$ transition, are compared
with previous limits given by antiprotonic Helium spectroscopy. Independent
constraints are also determined by investigating the effects of gravitational
spin-orbit coupling on the atomic spectrum. We show that the analysis of the
influence of that interaction, which is responsible for the spin precession
phenomena, on the fine structure of the states can be employed as a test of a
post-Newtonian potential in the atomic domain. The constraints obtained here
from 2P$_{1/2}-2P_{3/2}$ transition in hydrogen are tighter than previous
bounds determined from measurements of the spin precession in an
electron-nucleus scattering.

\end{abstract}
\maketitle

\section{Introduction}

The existence of extra dimensions has been speculated on a modern scientific
basis since the advent of the Kaluza-Klein theory in the early twentieth
century. This theory was an attempt to unify gravity and electromagnetism, the
interactions known at that time, within a single formalism that postulated the
presence of an additional spatial dimension. To avoid empirical
contradictions, it is assumed that the fifth dimension has the topology of a
circle with a radius of the order of the Planck length (10$^{-35}%
\operatorname{m}%
$). With such a tiny radius, however, there were no prospects for
experimentally testing the existence of hidden dimensions neither at that time
nor in a foreseeable future.

At the end of the last century, the subject of extra dimensions has emerged
with renewed interest due to the braneworld models, which proposes a new
scenario, inspired from developments in string theory, in which matter and all
the standard model fields are confined in a space with three spatial
dimensions (the 3-brane), while the gravitational field can propagate in all
directions \cite{ADD1,ADD2,RS1,RS2}.

Initially, this scenario drew a lot of attention because the predicted
"spreading" of gravity to other dimensions could be an explanation for the
hierarchy problem, i.e., for the question of why the gravitational interaction
is so weak compared to the other forces, at long distances
\cite{ADD1,ADD2,RS1,RS2}.

Another very interesting feature brought by some braneworld models, such as
the ADD model \cite{ADD1,ADD2}, is the possibility that empirical signals of
extra dimensions could be the object of experimental search at present days.
Indeed, as gravity is the only field that has access to the extra dimensions,
then the hypothesis that the compactification radius $R$ could be much greater
than the Planck length scale is phenomenological feasible, since gravity is
being tested in a small length scale just recently.

It is well known that, if the supplementary space has compact topology, the
gravitational field recovers its traditional four-dimensional behavior for
distances $r$ much greater than the compactification radius $R$. On the other
hand, the force gets strengthened at short distances ($r<R$). Direct
laboratory tests of the inverse-square law, based on modern versions of
torsion balances, put the upper limit $R<44$ $%
\operatorname{\mu m}%
$ \cite{Hoyle2007,review} if there is only one extra dimension. Regarding more
codimensions, the most stringent constraints come from high-energy particle
collision data and analysis of some astrophysics processes
\cite{SN,neutronstar,colliders,lhc,monojet,landsberg,pdgAlex}. In all cases,
however, experimental upper bounds for $R$ are much weaker than the Planck length.

The amplification of gravity at short distances as predicted by
higher-dimension theories has motivated many investigations about the behavior
of the gravitational field in microscopic domains
\cite{specforxdim1,atomicspec,Li,Wang,specforxdim2,molecule,dahia,fractal}. It
is worthy of mention that, concerning the proton radius puzzle
\cite{nature,science,carl,krauth,onofrio}, there are conjectures on the
possibility that higher-dimensional gravity could be an explanation for this
issue \cite{Dahia2}.

A very common way of expressing modifications of gravity is through the
so-called Yukawa parametrization, in which a Yukawa-like term is added to the
Newtonian potential \cite{review}. In this parametrization, the modified
gravitational potential is written as $\varphi=-GM/r\left(  1+\alpha
\exp(-r/\lambda\right)  )$, where the dimensionless parameter $\alpha$
measures the amplification of the interaction strength and $\lambda$
determines the length scale where the modifications are significant.

The Yukawa parametrization is very useful because it can account for
deviations of Newtonian gravity which may have different physics origins. As
we have mentioned above, hidden dimensions is a possible cause, however, there
are extensions of the standard model of particle physics that predict the
existence of additional bosons that indirectly could interfere in the inverse
square law of gravity in certain domains \cite{extensions,models}. Some
F(R)-theories make similar predictions too \cite{stelle}.

Facing these theoretical possibilities, here we are interested in using recent
spectroscopic data of the hydrogen in order to searching for modifications of
the gravitational field on the atomic scale. One of the tightest constraints
for the parameter $\alpha$, derived from the atomic spectroscopy, is imposed
by measurements of transition frequencies of the antiprotonic Helium
\cite{review,pHebound,5force}, by exploring the gravitational interaction
between the antiproton and the Helium nucleus. In section II, we shall see
that new data of the $1S-3S$ transition in the hydrogen establishes empiric
bounds on deviations of the Newtonian potential that are slightly stronger
than those obtained from that exotic atom.

Besides spectroscopic constraints, other independent bounds for hypothetical
short-range interactions around the Angstrom length scale can be inferred from
experiments that examine different physics phenomena such as neutron
scattering \cite{neutrons}, for instance. Another interesting example is the
MTV-G experiment \cite{mtvg,mtvg1} that intends to test the existence of a
strong gravitational field produced by atomic nuclei by measuring, with a Mott
polarimeter, the spin precession of an electron in a scattering process with a
heavy nucleus. Preliminary results were obtained by treating the geodetic
precession of the spin from a classical point of view \cite{review,mtvg,mtvg1}.

Here, inspired by the MTV-G experiment, we intend to inspect gravitational
effects on the spin precession of an electron which is found in an atomic
bound state, adopting the quantum perspective. As it is known, in the
Hamiltonian formalism, the geodetic spin precession is dictated by the
gravitational spin-orbit coupling \cite{fish,so}. In section III, considering
the Dirac equation in the curved spacetime, we discuss the influence of the
gravitational spin-orbit coupling in the fine splitting between the states
$2P_{1/2}$ and $2P_{3/2}$. As we shall see, this analysis does not put
empirical limits on deviations of the Newtonian potential only, but actually
it allows us to investigate the influence of a Post-Newtonian potential
associated with spatial components of the metric in atomic length scale. This
new potential is related to a specific parameter of the PPN-formalism
(parametrized Post-Newtonian formalism) \cite{ppn}, whose geometrical meaning
is connected to the curvature of the sections $t=const$. The constraints
obtained here are stronger than those derived from the MTV-G scattering
experiment \cite{review,mtvg}.

\section{Spectroscopic constraints}

In a certain range of length close to the Angstrom scale, the strongest
empirical constraint on deviations of the gravitational field, imposed by
atomic spectroscopy, as far as we know, comes from the analysis of transitions
in the antiprotonic Helium $(\bar{p}He^{+})$ \cite{review,pHebound,5force}.
This exotic atom is formed in laboratory by replacing one of the two electrons
of a natural Helium by an antiproton \cite{pHeDiscovery}. With this change,
the Coulombian interaction between the particles and the nucleus is not
directly altered, but the gravitational interaction between the antiproton,
with mass $m_{p}$, and the Helium's nucleus becomes almost two thousand times
bigger than that between the nucleus and the electron, since $m_{p}\simeq1836$
$m$, where $m$ is the electron mass. Hence, this exotic atom seems to be a
system with adequate features to investigate the behavior of gravity in the
atomic domain.

At first sight, another positive characteristic of this system would be the
possibility of probing the gravitational interaction in a range of length that
could reach thousandths of the Angstrom, once the relative distance between
the nucleus and the antiproton, which depends on the inverse of the antiproton
mass, could be much smaller than traditional Bohr radius $(a_{0}\simeq0,5%
\operatorname{\text{\AA}}%
)$. However, regarding this point, there is a downside aspect. In fact, in
states in which the antiproton is found very close to the nucleus, the
matter-antimatter annihilation process abbreviates the lifetime of the
antiprotonic Helium to few picoseconds \cite{pHe}, preventing, therefore, any
possibility of studying the spectroscopy of this atom with present technology.

It happens that, in a fraction of the antiprotonic helium atoms that are
synthesized in laboratory, $\bar{p}$ is found in Rydberg states with high
principal quantum number $n$ and high angular momentum $l\sim n-1$ \cite{pHe}.
In states with $n\sim l\sim40$, the average distance of the antiproton to the
nucleus is approximately around $a_{0}$ \cite{5force} and the overlap of the
wave function with the nucleus is drastically suppressed. As a consequence,
the lifetime of the antiprotonic helium increases to the order of microseconds
\cite{pHe}. These Rydberg states, with this longer lifetime, are amenable to
be investigated by laser spectroscopy and, indeed, the transition frequencies
between these metastable states were measured with a relative precision of
some parts in $10^{-9}$ \cite{pHeMeasure,pHe}. Comparing the experimental data
\cite{pHeMeasure,pHe} with the theoretical calculations
\cite{pHecalculations,pHeCalc2}, based on the theory of Quantum
Electrodynamics (QED), one verifies a precise agreement between them
\cite{review,pHebound,5force}. This result put some bounds on the values that
the parameters $\alpha$ and $\lambda$ could assume. For instance, at $1\sigma$
confidence-level, $\alpha<10^{28}$ for $\lambda\sim1%
\operatorname{\text{\AA}}%
$ \cite{pHebound}.

Here, we investigate possible deviations of the gravitational Newtonian
potential in the atomic scale by using new data from hydrogen spectroscopy.
The intention is to compare the constraints for the Yukawa parameters
determined by the hydrogen spectroscopy with that established by the
antiprotonic Helium. Although the gravitational interaction between proton and
electron in the hydrogen atom is much weaker than the antiproton-nucleus
gravitational interaction in $\bar{p}He^{+}$, the available spectroscopic data
of the hydrogen are much more accurate. For instance, the experimental value
of the $1S-2S$ transition frequency, $f_{1S-2S}^{\exp}=2466061413187035$ Hz,
was recently measured with an error of $\delta_{\exp}=10$ Hz, which
corresponds to a relative precision of the order of $10^{-14}$ \cite{H}. If
the theoretical value $f_{1S-2S}^{th}$, predicted by QED, had an uncertainty
$\delta_{th}$ of the same magnitude order, then the agreement between
$f_{1S-2S}^{th}$ and $f_{1S-2S}^{\exp}$ would impose a much tighter constraint
for the Yukawa parameters (see Figure 1). However the experimental value of
the $1S-2S$ transition frequency is the most precise value of the data set
that is employed to determine the values of certain fundamental spectroscopic
constants, such as the Rydberg constant (see Table XVIII of Ref.
\cite{CODATA2010}). As the theoretical predictions depend on these constants,
the comparison between the calculated value with the measured frequency of
this specific transition should be viewed with caution \cite{3s1sCalc}.

Because of this, let us consider the $1S-3S$ transition. The relative
precision achieved in the most recent measurement of the frequency of this
transition, $f_{1S-3S}^{\exp}$, is of the order of $10^{-12}$ \cite{3s1s}. The
theoretical value calculated in Ref. \cite{3s1sCalc} is of the same order.
Although $f_{1S-3S}^{\exp}$ is not so accurate as $f_{1S-2S}^{\exp}$, the
advantage of using that frequency to test QED is the fact that the isolated
value of $1S-3S$ transition frequency does not belong to the input data used
in the least-squares adjustment of the values of the fundamental constants
recommended by CODATA-2002\cite{CODATA2002}, which were employed by Ref.
\cite{3s1sCalc} to calculate the frequency of that transition.

The theoretical \cite{3s1sCalc} and experimental \cite{3s1s} values are
respectively:%
\begin{align}
f_{1S-3S}^{th}  &  =2922743278671.6(1.4)%
\operatorname{kHz}%
,\\
f_{1S-3S}^{\exp}  &  =2922743278671.5(2.6)%
\operatorname{kHz}%
.
\end{align}
The uncertainty is expressed in parenthesis. Admitting that the theoretical
and experimental uncertainties are independent, then the combined error is
$\delta f=\sqrt{\delta_{th}^{2}+\delta_{\exp}^{2}}\simeq3.0%
\operatorname{kHz}%
$. It is clear that the theoretical prediction $f_{1S-3S}^{th}$ agrees very
well with the measured frequency $f_{1S-3S}^{\exp}$ within the combined error
$\delta f$. Therefore, any new hypothetical interaction, such as the modified
gravitational interaction, should not introduce corrections for the transition
frequency in an amount $\Delta f$ greater than the error $\delta f$. The
condition $\Delta f<\delta f$ imposes certain empirical limits for the
parameters $\alpha$ and $\lambda$.

The supposed correction, $\Delta f$, that the modified gravity provides for
the $1S-3S$ transition can be calculated by using the perturbation method. The
new gravitational interaction between the proton and electron is described, in
the leading order, by the Hamiltonian $H_{G}^{(0)}=m_{e}\varphi$, where
$\varphi=-Gm_{p}/r\left(  1+\alpha\exp(-r/\lambda\right)  )$ is the modified
gravitational potential produced by the proton. This Hamiltonian $H_{G}^{(0)}$
should be considered as a small term of the atomic Hamiltonian. According to
the perturbation formalism, in the first order, this new interaction will
decrease the energy of each state $\psi$ by the amount $\left\langle
H_{G}^{(0)}\right\rangle $, the average value of $H_{G}^{(0)}$ in the state
$\psi$. The gravitational interaction will change the energy of the states
$1S$ and $3S$ by different amounts, increasing the energy gap between these
states. This implies a correction of the transition frequency which, in the
first approximation order, is given by:
\begin{equation}
\Delta f=\frac{\left\langle H_{G}^{(0)}\right\rangle _{3S}-\left\langle
H_{G}^{(0)}\right\rangle _{1S}}{h}. \label{f}%
\end{equation}
\qquad\qquad\qquad\qquad

In Figure 1, we show the constraints on the Yukawa parameters imposed by the
condition $\Delta f<\delta f.$ As we can see, the new bounds are slightly
stronger than those obtained from the spectroscopy of the atom $\bar{p}He^{+}%
$. For $\lambda=1%
\operatorname{\text{\AA}}%
$, for instance, the data demand that $\alpha<1.7\times10^{27}$ at $1\sigma$
confidence-level.%
\begin{center}
\includegraphics[
height=2.1439in,
width=3.3122in
]%
{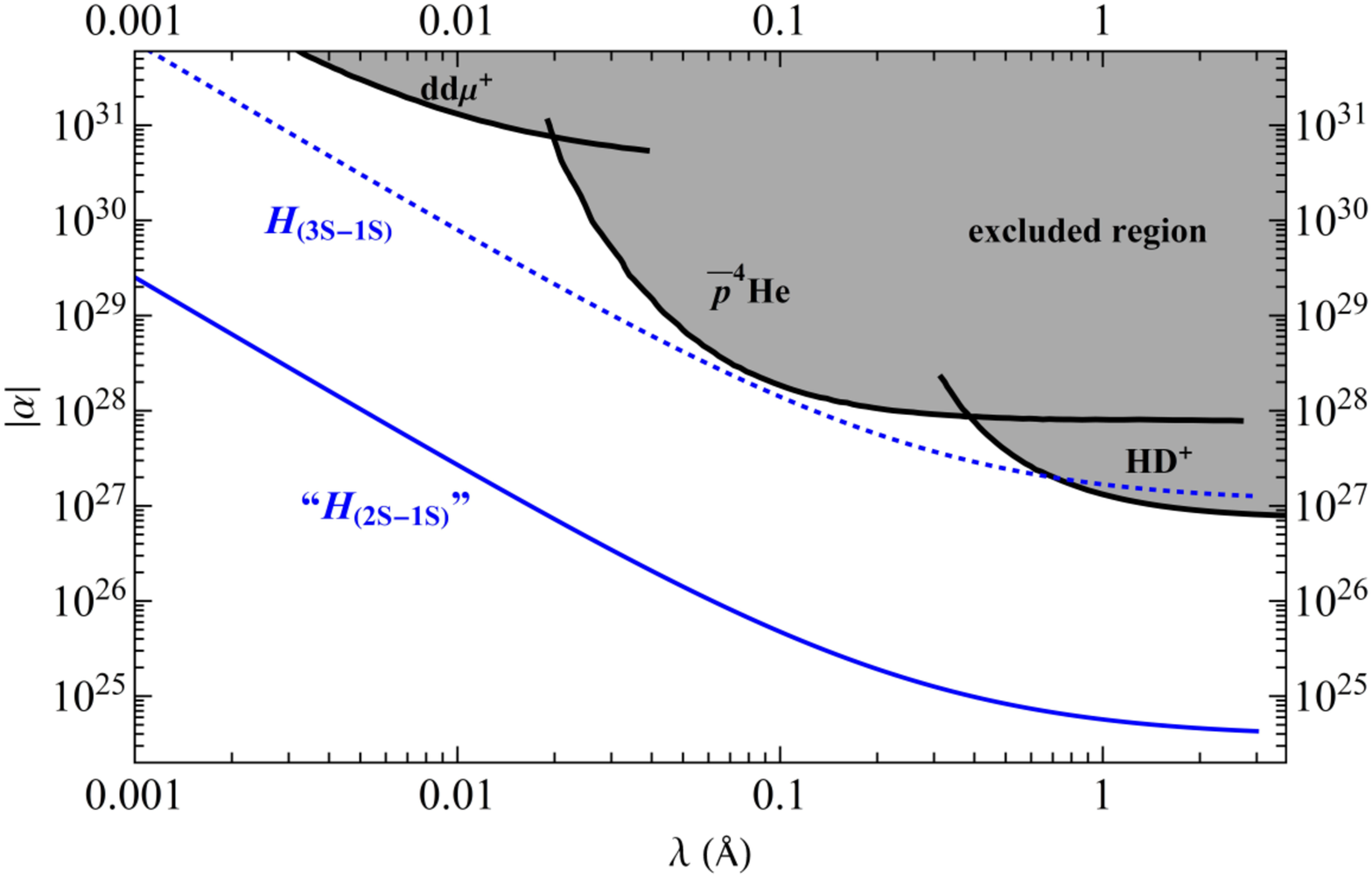}%
\\
Figure 1. \textit{Constraints for deviations of Newtonian potential determined
here from the }$1S-3S$\textit{ transition in the hydrogen (}$H_{1S-3S}%
$\textit{ line) are compared to other bounds imposed by spectroscopy of
}$dd\mu^{+},$\textit{ }$\bar{p}He^{+}$\textit{ and }$HD^{+}$\textit{(data
extracted from Ref. \cite{5force}). The }$H_{1S-2S}$\textit{ curve is just a
reference line that shows how stringent the spectroscopy of hydrogen could be,
considering the current relative empirical precision (10}$^{-14}$\textit{) of
the }$1S-2S$\textit{ transition.}%
\label{Fig1}%
\end{center}

Actually, according to Figure 1, for $\lambda<$ $0.6%
\operatorname{\text{\AA}}%
$, the $H_{1S-3S}$ constraints are tighter than several spectroscopic bounds,
which also include empirical limits determined by the spectroscopy of
$dd\mu^{+}$ (an exotic molecule formed by two deuterons and one muon) and also
from the HD$^{+}$ (the ionized molecule constituted by a hydrogen and a deuterium).

The frequency of the $1S-3S$ transition that we use here is one of the most
accurate measurements in hydrogen spectroscopy, only surpassed by the $1S-2S$
transition \cite{3s1s proton}. Based on these two transitions, it is possible
to infer the proton charge radius. The value found in Ref. \cite{3s1s proton}
is in agreement with the CODATA-2014 recommended value and differs by
$2.8\sigma$ from the value extracted from the muonic hydrogen spectroscopy.
Given this result, it seems interesting to resort to Rydberg states if we are
aiming for a spectroscopic analysis that is less dependent on the proton size
\cite{Jentschura}.

The effects of hidden dimensions in certain Rydberg states were studied in
Ref. \cite{dahia3}, by using a power-law parametrization for the modified
gravitational potential. More recently, in Ref. \cite{Ryd} (we thank one of
the referees to call our attention to this paper) Rydberg states were also
considered with the purpose of constraining non-standard interactions by using
Yukawa parametrization. As expected, the strongest restrictions are found in a
length scale beyond the Bohr radius, since the studied states have a large
principal quantum number. Admitting a relative precision of the order of
$10^{-12}$ in the energy levels, it was found that the strength of the new
interaction should be lesser than $10^{28}$ for $\lambda>10^{-9}$m (after
converting data of figure 3 of Ref. \cite{Ryd} to units used here), which is
very close to our result. They also considered the Rydberg states combined
with data from other transitions. In this case, the constraint for $\alpha$ is
almost of the order of $10^{27}$ in the range $10^{-10}-10^{-7}%
\operatorname{m}%
$ at a 95\% confidence level, which is clearly compatible with our result.

So far, we have considered only spectroscopic constraints, since the main
objective of this work is to use new data from hydrogen spectroscopy to put
indenpendent constraints on deviations of standard gravity. However, it is
also interesting to compare these constraints with empirical limits imposed by
sources of different natures. Figure 2, in addition to the spectroscopic
constraints, includes other empirical limits set by data with distinct origins
such as particle colliders, Casimir effect, torsion balance and Lunar Laser
Ranging experiment, among others.%

\begin{center}
\includegraphics[
height=2.1439in,
width=3.333in
]%
{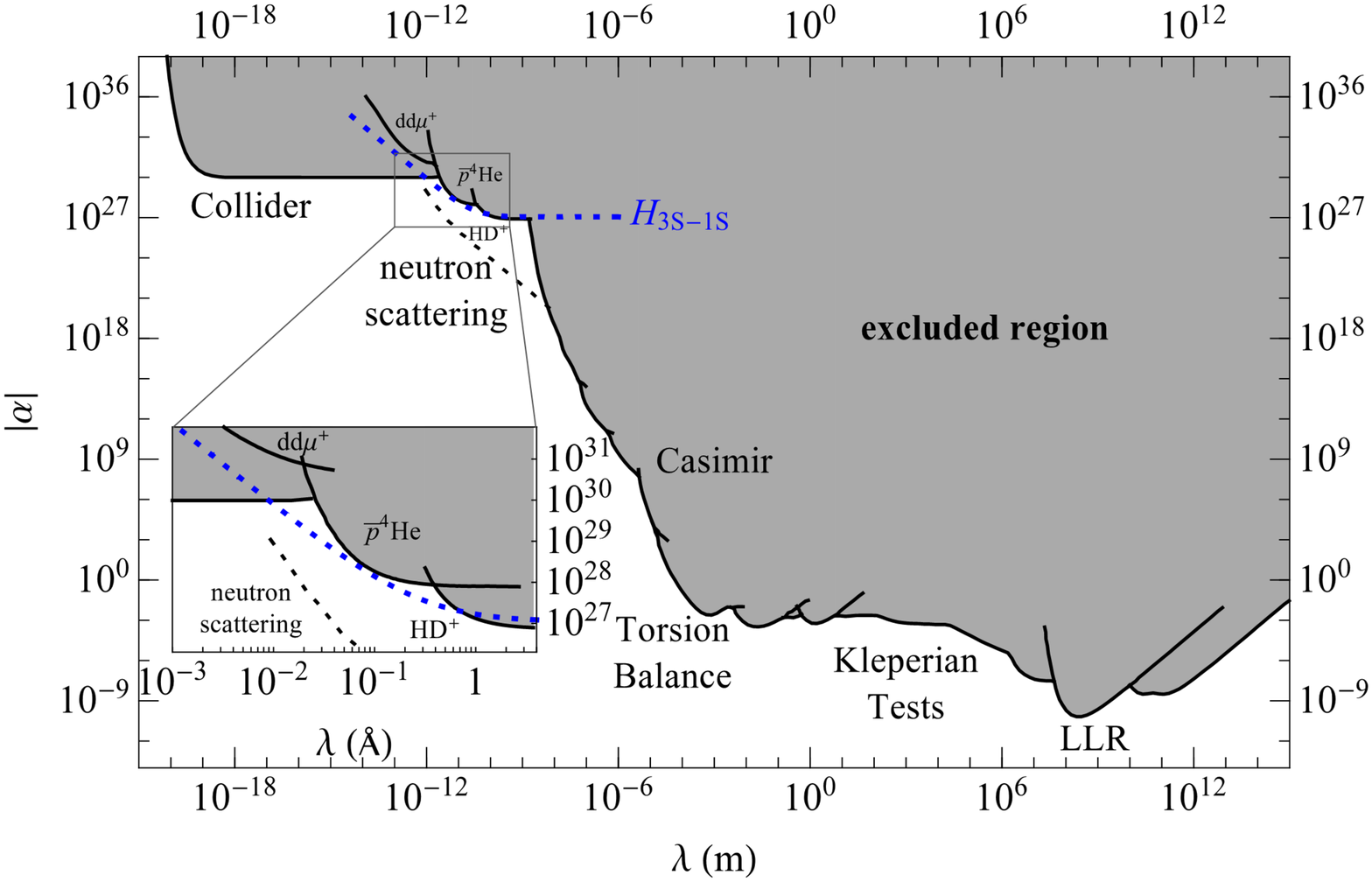}%
\\
Figure 2. \textit{In this figure, the spectroscopic constraints for }$\alpha$
\textit{determined here (H}$_{1S-3S}$\textit{ line) are compared to empirical
limits established by data from different origins (see Ref. \cite{review} for
collider data and Ref. \cite{5force} for all other data).}%
\label{Figure2}%
\end{center}

Almost all data in Figure 2 were extracted from Ref. \cite{5force}, except for
the collider data, which are obtained from Figure 1 of Ref. \cite{review}, and
the line H$_{3S-1S}$ which was determined here. Each distinct bound stands out
on different length scale. As we can see, in this more general context, the
$H_{1S-3S}$ constraint is surpassed by the collisor bound for short $\lambda$
and by HD$^{+}$ limit for $\lambda>0.6$ $%
\operatorname{\text{\AA}}%
$. It is also important to remark that constraints from neutron scattering,
claimed in Ref. \cite{neutrons}, are stronger than spectroscopic limit.

\section{Gravitational spin-orbit coupling}

In a curved space, the final direction of a vector that is parallel
transported along a closed path may not coincide with its initial direction.
Based on this fact, it is expected, in accordance with the theory of General
Relativity, that when a particle moves in a gravitational field its spin will
precess \cite{schiff}.

This effect in the classical regime was directly verified by some experiments
such as GRAVITY PROBE B \cite{probeB}, which measured the orientation changes
of a gyroscope's axis moving along a geodesic around the Earth.

In the microscopic domain, the precession of an electron's spin induced by the
gravitational field of an atomic nucleus was preliminarily investigated in the
so-called MTV-G experiment \cite{mtvg,mtvg1}. This experiment tests the
existence of a strong gravitational field in the nuclear domain by studying
the influence of a modified gravity on the spin precession of an electron that
is scattered by a nucleus. Following a classical description of the geodetic
spin precession of the electron, some preliminary results were obtained
\cite{review,mtvg,mtvg1}.

However, as the spin of an elementary particle is a quantum quantity, it is
more appropriate to treat the spin precession phenomena induced by a
gravitational field in the quantum mechanics formalism.

For this purpose, let us consider the Dirac equation in curved spacetime. The
coupling between fermions and the gravitational field is implemented through
the tetrad fields $e_{\hat{A}}^{\mu},$ which consists of components of four
orthonormal vector fields (each one is identified by the index $\hat{A}=0,1,2$
or $3$) written in some coordinate system $\left(  x^{\mu}\right)  $ (here,
$\mu=0,1,2$ and $3$). The tetrad fields satisfy the orthonormality condition
$g_{\mu\nu}e_{\hat{A}}^{\mu}e_{\hat{B}}^{v}=\eta_{\hat{A}\hat{B}}$ , where
$g_{\mu\nu}$ denotes the metric of the spacetime and $\eta_{AB}$ is the
Minkowski metric.

With the help of the tetrad fields and the Dirac matrices defined on the
Minkowski spacetime, $\gamma^{\hat{A}}$, we can construct the matrices
$\gamma^{\mu}\left(  x\right)  =e_{\hat{A}}^{\mu}\left(  x\right)
\gamma^{\hat{A}}$, which satisfy the anti-commutation algebra of the Dirac
matrices adapted to the curved space: $\left\{  \gamma^{\mu}\left(  x\right)
,\gamma^{\nu}\left(  x\right)  \right\}  =2g^{\mu\nu}\left(  x\right)  $.

The Dirac equation that describes the state $\psi$ of a particle of mass $m$
in a gravitational field is given by%
\begin{equation}
\left[  i\gamma^{\mu}\left(  x\right)  \nabla_{\mu}-mc/\hbar\right]
\psi\left(  x\right)  =0,
\end{equation}
where $c$ is the speed of light, $\hbar$ is the reduced Planck constant
and\emph{ }$\nabla_{\mu}$ is the covariant derivative of the spinor $\psi$,
which depends on the spinorial connection $\Gamma_{\mu}\left(  x\right)  $ as
follows:%
\begin{equation}
\nabla_{\mu}\psi\left(  x\right)  =\left[  \partial_{\mu}+\Gamma_{\mu}\left(
x\right)  \right]  \psi\left(  x\right)  . \label{MP02}%
\end{equation}
Admitting the compatibility between the spinorial connection and the metric,
it is possible to write $\Gamma_{\mu}\left(  x\right)  $ in terms of the
Levi-Civita covariant derivative of the tetrad fields, $e_{\hat{A};\mu}%
^{v}\left(  x\right)  $, according to the expression:
\begin{equation}
\Gamma_{\mu}\left(  x\right)  =-\frac{i}{4}\sigma^{\hat{A}\hat{B}}g_{\alpha
\nu}e_{\hat{A}}^{\alpha}\left(  x\right)  e_{\hat{B};\mu}^{\nu}\left(
x\right)  , \label{MP03}%
\end{equation}
where $\sigma^{\hat{A}\hat{B}}=\frac{i}{2}\left[  \gamma^{\hat{A}}%
,\gamma^{\hat{B}}\right]  $ is a representation of the Lorentz Lie Algebra in
the spinor space, written in terms of the commutating operator $[,]$.

From the Dirac equation, we can study the influence of the gravitational field
produced by the proton on the behavior of an electron in the hydrogen atom. In
the first approximation approach, it is reasonable to assume that the proton
produces a static gravitational field with spherical symmetry. Under this
condition, the spacetime metric can be put in the following form:%
\begin{equation}
ds^{2}=-c^{2}w^{2}dt^{2}+v^{2}(dx^{2}+dy^{2}+dz^{2}), \label{metric}%
\end{equation}
in the isotropic coordinates. The functions $w$ and $v$ depend only on the
coordinate $r=\left(  x^{2}+y^{2}+z^{2}\right)  ^{1/2}$. Associated to this
metric, the non-null tetrad fields components are:
\begin{align}
e_{\hat{0}}^{0}\left(  x\right)   &  =w^{-1},\label{e0}\\
e_{\hat{\jmath}}^{i}\left(  x\right)   &  =\delta_{j}^{i}v^{-1}. \label{e1}%
\end{align}
In the weak-field regime the function $v$ and $w$ can be expressed in terms of
gravitational potentials as:%
\begin{align}
w  &  =1+\varphi/c^{2},\label{w}\\
v  &  =1-\tilde{\varphi}/c^{2}. \label{v}%
\end{align}
According to the General Relativity theory, $\varphi=\tilde{\varphi}$.
However, as we are\emph{ }investigating modifications of the gravitational
field in the atomic domain, let us admit that $\varphi$ and $\tilde{\varphi}$
can be different functions, or more precisely, that the Yukawa parameter
($\tilde{\alpha}$) associated to the potential $\tilde{\varphi}$ is not
necessarily equal to $\alpha$ (the parameter investigated in the previous
section) and should be determined experimentally.

The potential $\tilde{\varphi}$ is directly related to the curvature of the
spatial section $(t=const.)$ of the spacetime according to the geometric
viewpoint. The possibility that $\tilde{\varphi}$ is not necessarily equal to
the Newtonian potential is also embodied in the parametrized post-Newtonian
(PPN) formalism, which is a theoretical framework properly developed for the
purpose of testing metric theories, such as the General Relativity and
Brans-Dicke theory, in the weak-field limit \cite{ppn}.

The parameter $\tilde{\alpha}$ that we are investigating here can be put in
correspondence with a certain PPN-parameter. Considering the Yukawa
parametrization of the potential, we can see that $\tilde{\varphi
}=GM/r(1+\tilde{\alpha})$, in the limit $r<<\lambda$. Thus, we can conclude
that the combination $\left(  \tilde{\alpha}+1\right)  $ plays the role of the
parameter $\gamma$ of the PPN-formalism \cite{ppn}.

In the astrophysics domain, empirical bounds for the PPN-parameter $\gamma$
are extracted from time-delay and light deflection experiments, for example.
Recent experiments were performed with the help of Cassini spacecraft and find
that $\gamma=1+(2.1\pm2.3)\times10^{-5}$ \cite{gamma}, by studying the
behavior of radio waves under the influence of the gravitational field of the
Sun. This constraint is valid in the length scale of the solar radius and is
compatible with the value predicted by the theory of General Relativity
$\left(  \gamma=1\right)  $.

As we shall see later in this section, this Post-Newtonian parameter can be
investigated in the atomic domain through the study of the influence of the
gravitational spin-orbit coupling on the fine structure of the atom.

To show this, let us turn our attention to the Dirac equation again, now using
the tetrad fields given above (Eqs. (\ref{e0}) and (\ref{e1}) ). The Dirac
equation can be rewritten in the form $i\hbar\frac{\partial\psi}{\partial
t}=H_{G}\psi$, where $H_{G}$ is the operator that contains the gravitational
sector of the atomic Hamiltonian. In a convenient representation, $H_{G}$
assumes the following form in the first order of the gravitational potentials
\cite{fish,so,DiracSchwarz}:%
\begin{equation}
H_{G}=\beta mc^{2}+\beta\varphi m+\frac{1}{2}\left\{  \vec{\alpha}\cdot\vec
{p},[1+(\varphi+\tilde{\varphi})/c^{2}]\right\}  , \label{H}%
\end{equation}
where $\vec{p}$ is the usual three-dimensional momentum operator in flat
spacetime, $\alpha^{i}=$ $\gamma^{0}\gamma^{i}$ and $\beta=\gamma^{0}$. The
first term of $H_{G}$ is related to the rest energy of the electron, the
second term is associated with the usual potential energy of the
proton-electron gravitational interaction ($H_{G}^{(0)}$), which was examined
in the last section, and the third one gives rise to the kinetic term and also
to relativistic and quantum corrections. In the form (\ref{H}), it is
important to stress that $H_{G}$ is Hermitian in the Hilbert space of
square-integrable functions endowed with the usual flat inner product ($\int
d^{3}x$).

Admitting that the rest energy of the test particle is the leading term of the
Hamiltonian, a semi-relativistic expansion of $H_{G}$ (\ref{H}) can be
obtained by following the Foldy-Wouthuysen procedure, which consists in the
elimination of odd operators of $H_{G}$, in each order of $1/mc^{2}$, by a
convenient sequence of unitary transformations of the Hamiltonian \cite{FW}.
Among many terms that arise in the expansion, here we want to focus our
attention on the Hamiltonian term associated with the gravitational spin-orbit
coupling, which can be expressed as \cite{fish,so}:
\begin{equation}
H_{Gso}=\frac{1}{mc^{2}}\frac{1}{r}\left(  \frac{1}{2}\frac{d\varphi}%
{dr}+\frac{d\tilde{\varphi}}{dr}\right)  (\vec{S}\cdot\vec{L}), \label{HGso}%
\end{equation}
where $\vec{L}$ is the orbital angular momentum of the particle and $\vec{S}$
is the spin operator, which can be written in terms of the Pauli matrices in
the form $\vec{S}=(\hbar/2)\vec{\sigma}$.

As we have already mentioned, this term (\ref{HGso}), in the classical regime,
is responsible for the geodetic precession of the axis of a gyroscope in
curved spacetime \cite{fish,so,schiff}. Considering $\vec{S}$ as a classical
angular momentum measured by a co-moving geodesic observer, it can be shown
\cite{so,schiff} that the spin dynamics is governed by the equation:%
\begin{equation}
\frac{d\vec{S}}{dt}=\left[  \left(  \frac{1}{2}+\gamma\right)  \frac
{GM}{mc^{2}r^{3}}\vec{L}\right]  \times\vec{S}, \label{spin}%
\end{equation}
assuming that $\varphi=\tilde{\varphi}/\gamma=-GM/r$.

The analysis of the spin precession in the MTV-G experiment was based on the
equation (\ref{spin}) taking $\gamma=1$ \cite{mtvg,mtvg1}. Therefore, without
making any distinction between the Newtonian and the post-Newtonian potentials.

In the present work, we want to study the effect of the gravitational
spin-orbit coupling on the energy levels of the hydrogen. In the atom, the
electron is not a free-falling particle, but it is found in a bound state due
to the electromagnetic interaction with the proton. This dominant interaction
establishes a spin-orbit coupling too, described by a Hamiltonian that can be
put in the form \cite{FW}:%

\begin{equation}
H_{Eso}=\frac{-q}{2m^{2}c^{2}}\frac{1}{r}\frac{d\phi_{E}}{dr}\left(  \vec
{S}\cdot\vec{L}\right)  , \label{HEso}%
\end{equation}
where $-q$ is the electron charge and the function $\phi_{E}$ is the electric
potential produced by the proton in the curved space. In a spacetime with the
metric (\ref{metric}), the electric field equations can be written in the same
form of the Maxwell equations defined in a flat space endowed with a new
electric permittivity given by $\varepsilon=\varepsilon_{0}v/w$, where
$\varepsilon_{0}$ is the permittivity of free space \cite{fish,dahia3}. Thus,
if the electrostatic potential has spherical symmetry, it satisfies, in the
first-order approximation, the following equation:
\begin{equation}
\frac{d\phi_{E}}{dr}=-\frac{q}{4\pi\varepsilon_{0}r^{2}}\left(  1+\varphi
/c^{2}+\tilde{\varphi}/c^{2}\right)  . \label{phiE}%
\end{equation}
As we can see, the spacetime curvature changes the electric potential. Through
this correction of $\phi_{E}$, the gravitational field can indirectly
influence the atomic spin-orbit coupling from Hamiltonian $H_{Eso}$. However,
as we show in appendix, this contribution is five magnitude orders (10$^{-5}$)
lesser than the direct contribution coming from (\ref{HGso}). So, for our
purposes, the indirect gravitational contribution can be ignored hereafter.

It is well known that the electromagnetic spin-orbit coupling is responsible
for a fine splitting of energy levels with the same angular momentum $l$, such
as the $2P_{1/2}$ and $2P_{3/2}$ states, for example. The gravitational
analogous (\ref{HGso}), considered here as a weaker interaction in comparison
to the Hamiltonian $H_{Eso}$, will provide an additional shift in those states.

There are precise theoretical calculations of the energy of the hydrogen
states, based on the QED theory. In Ref. \cite{CODATA2010}, aiming to test the
QED predictions, it was explicitly determined the transitions frequency
between levels with $n=2$ in hydrogen. Specifically, the frequency transition
between $2P_{1/2}$ and $2P_{3/2}$ is (pp. 1540 of Ref. \cite{CODATA2010}):%
\begin{equation}
f_{2P_{1/2}-2P_{3/2}}^{th}=10969041.571(41)%
\operatorname{kHz}%
.
\end{equation}
This calculation is based on the recommended values for the fundamental
constants which are extracted from the input data of CODATA-2010, but
excluding the experimental values of $2S_{1/2}-2P_{1/2}$ and $2P_{3/2}%
-2S_{1/2}$ transition frequencies, which corresponds to the items A39, A40.1,
and A40.2 in Table XVIII of CODATA-2010 \cite{CODATA2010}.

On its turn, there are measurements of the centroid transition frequencies
between $2S_{1/2}-2P_{1/2}$ \cite{2s2p12} and $2P_{3/2}-2S_{1/2}$
\cite{2s2p32}. Based on these experimental values, from which the hyperfine
structure of the states has already been excluded, we can determine the
experimental frequency of the transition between $2P_{1/2}$ and $2P_{3/2}:$%
\begin{equation}
f_{2P_{1/2}-2P_{3/2}}^{\exp}=10969045(15)%
\operatorname{kHz}%
.
\end{equation}
Of course, the theoretical and experimental values coincide within the
combined error $\delta f=$ 15 $%
\operatorname{kHz}%
$. So the contribution provided by the gravitational spin-orbit interaction,
$\Delta f_{so}$, cannot exceed this empirical error.

Now, in order to estimate $\Delta f_{so}$, let us remember that, due to the
spin-orbit interaction, the operators $\vec{L}$ and $\vec{S}$ no longer
commute with the atomic Hamiltonian. On the other hand, the total angular
momentum $\vec{J}=\vec{L}+\vec{S}$ is a conserved quantity. Therefore, atomic
stationary states are labeled by the eigenvalues of the total angular
momentum. When the orbital angular momentum and the spin are, let us say,
aligned, the total momentum is $j=l+1/2$ and, in this case, $H_{Gso}$ will
provide a positive energy shift for the state. On the other hand, for states
with $j=l-1/2$, the spin-orbit interaction will give a negative contribution
to the energy.

These effects increase the energy gap between those states. In the first
approximation, the additional separation between states $(n,l,j=l+1/2)$ and
$(n,l,j=l-1/2)$ due to gravitational spin-orbit interaction leads to the
following change in the frequency transition:%
\begin{equation}
\Delta f_{so}=\frac{\Delta E_{Gso}}{h}=\frac{1}{hmc^{2}}\left\langle \frac
{1}{r}\frac{d}{dr}(\varphi/2+\tilde{\varphi})\right\rangle _{n,l}\left(
l+1/2\right)  \hbar^{2}, \label{fso}%
\end{equation}
where the average $\left\langle {}\right\rangle _{n,l}$ is calculated with
respect to the radial solution of the Schr\"{o}dinger equation, $R_{n,l}%
\left(  r\right)  $, for states with a principal quantum number $n$ and
orbital angular momentum $l$.

Now let us discuss the implications of the condition $\Delta f_{so}<\delta f$
in the case of the $2P_{1/2}-2P_{3/2}$ transition, for which $R_{21}=\left(
1/\sqrt{24a_{o}^{3}}\right)  (r/a_{o})e^{-r/2a_{0}}$. First, let us emphasize
that in the previous section, we have tested modifications of the Newtonian
potential, but not the potential $\tilde{\varphi}$, since, in the $1S-3S$
transition, the curvature of the spatial sections has no influence on the
energy of the $S$-states in the leading order.

The expression (\ref{fso}) indicates that the effects of the gravitational
spin-orbit coupling on the spectroscopy depend on a linear combination of the
Newtonian potential and the potential $\tilde{\varphi}$. Therefore, the
spectroscopic analysis will put empirical bounds on the mixed parameter
$\left(  \alpha/2+\tilde{\alpha}\right)  $. However, taking into account the
empirical bounds for the parameter $\alpha$ previously established, we can
verify that the contribution of the potential $\varphi$ in the transition
$2P_{1/2}-2P_{3/2}$ is smaller than the experimental error. Therefore, for
practical purposes, we can neglect the potential $\varphi$ in the expression
(\ref{fso}). Thus, we can conclude that the analysis of the influence of the
gravitational spin-orbit interaction on the fine structure of the states is
actually a test of the Post-Newtonian potential $\tilde{\varphi}$ in the
atomic domain.

In Figure 3, we show the constraints for the mixed parameter $\left(
\alpha/2+\tilde{\alpha}\right)  $ in terms of $\lambda$. As we can see, for
$\lambda>1.5\times10^{-3}%
\operatorname{\text{\AA}}%
$, the bounds determined by the $2P_{1/2}-2P_{3/2}$ transition are more
stringent than the empirical limits put by MTV-G experiment. In particular,
for $\lambda=1$ $%
\operatorname{\text{\AA}}%
$, we find $\tilde{\alpha}<$ $2.1\times10^{33}$, which is stronger than the
MTV-G constraint by four magnitude orders.%

\begin{center}
\includegraphics[
height=2.1439in,
width=3.3131in
]%
{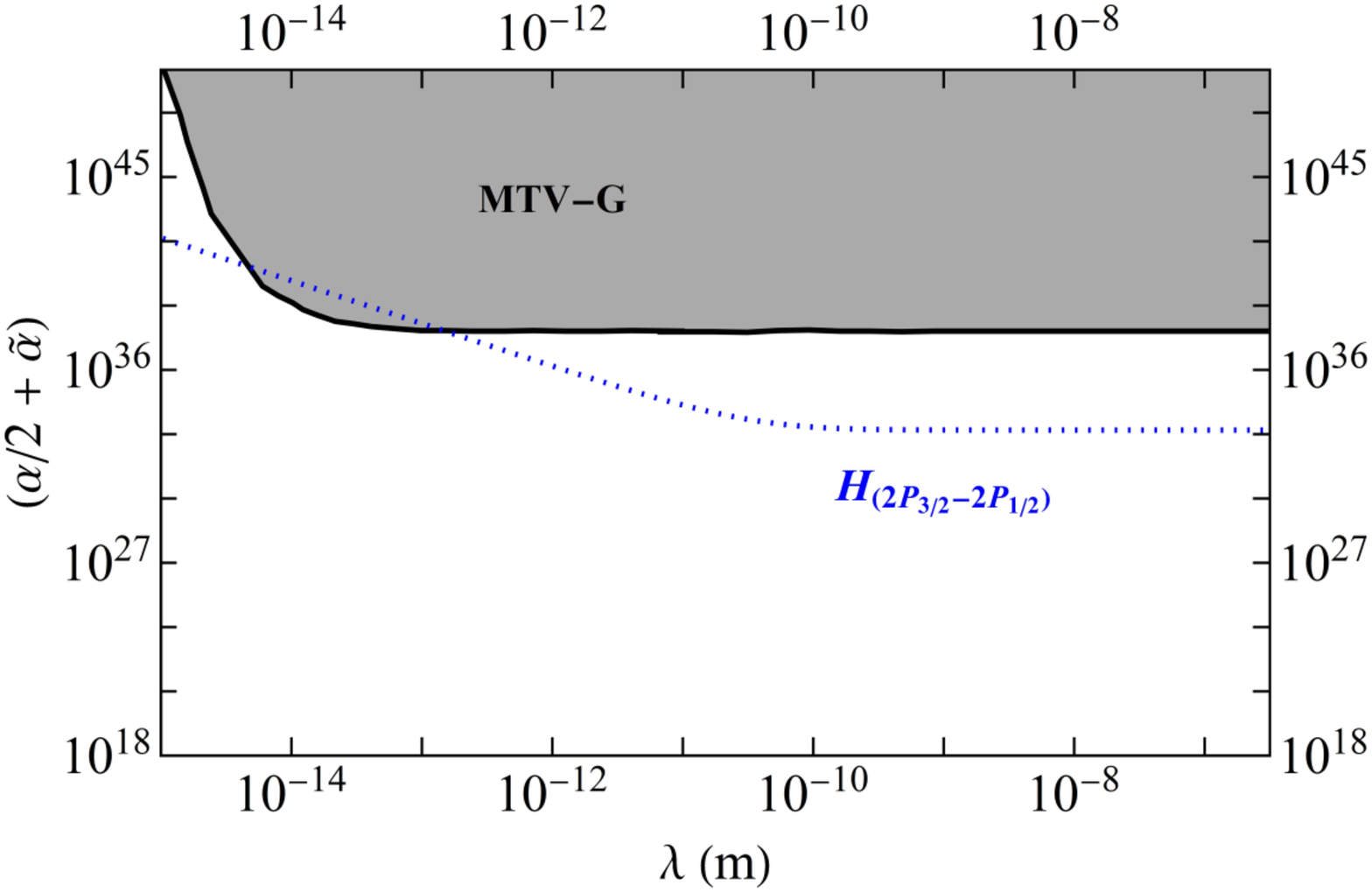}%
\\
\textit{Figure 3. The dashed line is extracted from the analysis of the
influence of the gravitational spin-orbit coupling on the separation between
the states }$2P_{1/2}$\textit{ and }$2P_{3/2}$\textit{ of hydrogen. It sets an
empirical constraint for the mixed parameter }$\left(  \alpha/2+\tilde{\alpha
}\right)  $\textit{, where }$\tilde{\alpha}$\textit{ is related to the
}$\gamma$\textit{-parameter in the PPN-formalism. For }$\lambda>1.5\times
10^{-3}\operatorname{\text{\AA}}$, \textit{the }$2P_{1/2}-2P_{3/2}$\textit{
constraint is stronger than the empirical bounds extracted from the MTV-G
scaterring experiment\cite{mtvg1}}.
\label{Fig3}%
\end{center}

The analysis based on the influence of the gravitational spin-orbit coupling
on the $2P_{1/2}-2P_{3/2}$ transition provides a weaker limit compared to the
bounds determined from the $1S-3S$ transition. However, we should have in mind
that these two tests probe different physical quantities. In fact, the $1S-3S$
transition, as well as all tests described in the Figure 2, sets upper limits
on deviation of the Newtonian potential, while the $2P_{1/2}-2P_{3/2}$
actually allows us to constrain the behavior of a post-Newtonian potential,
which has a proper geometric meaning and it is not necessarily equal to the
Newtonian potential in some metric theories.

\section{Final remarks}

Considering accurate data of hydrogen spectroscopy (more specifically, recent
measurement of the $1S-3S$ transition frequency \cite{3s1s}) we find
constraints for short-range modifications of the Newtonian potential in the
atomic domain. The bounds obtained here are tighter than several empirical
limits imposed by the spectroscopy of some exotic atom such as the
antiprotonic-Helium. Although the interaction between electron and proton is
almost two thousand times weaker compared to antiproton-nucleus gravitational
interaction, we have seen that the accuracy achieved in the hydrogen
spectroscopy is high enough to determine an improved constraint for deviations
of the Newtonian potential in range $\lambda<0.6%
\operatorname{\text{\AA}}%
$.

In the present discussion, we have adopted the comprehensive Yukawa
parametrization to express modifications of gravity. The reason is that
several models, such as large extra dimension models \cite{kehagias} and some
F(R)-theories \cite{stelle}, predict an exponentially decreasing correction of
the Newtonian potential in a domain beyond a certain length scale. Although
the Yukawa parametrization is useful, it has some limitations as any
approximation scheme\cite{newphys}. Indeed, as we are exploring atomic
spectroscopy data, the best results are found when the supposed deviations
occur around the Bohr radius ($a_{0}$). However, if the modifications are
significant in a length scale much lesser than $a_{0}$, then the Yukawa
parametrization will not be capable to appropriately capture their effects. In
this case, each model should be considered separately and studied in detail.
In Ref. \cite{dahia}, for instance, the power-law parametrization is adopted
to study the ADD model in thick branes scenarios.

In this paper, we have also investigated non-standard gravitational effects on
the fine separation between the $2P_{3/2}$ and $2P_{1/2}$ states. The energy
difference between these states, which have the same principal quantum number
and the same orbital angular momentum, is mainly determined by the spin-orbit
coupling. This interaction, which is responsible for the spin precession
phenomena, was explored by MTV-G experiment in order to investigate a possible
strong gravitational field produced by the nucleus, by measuring the spin
precession of an electron in a scattering process with a Mott polarimeter.

Inspired by the MTV-G experiment, we have discussed the effects of
gravitational spin-orbit coupling on the electron in a bound state. From the
Dirac equation, we study the influence of that interaction on the fine
structure of the $2P$-state of the hydrogen. We have shown that the empirical
constraints determined by this analysis are numerically weaker than those put
by the $3S-1S$ transition, however, it is important to emphasize that, as the
separation between $2P_{3/2}$ and $2P_{1/2}$ states has essentially a
relativistic origin, then the analysis of the $2P_{3/2}-2P_{1/2}$ transition
yields not a test of Newtonian potential, but, actually, it is a test of a
post-Newtonian potential, which is related to the $\gamma$-parameter in the PPN-formalism.

We have seen that, for $\lambda>1.5\times10^{-3}%
\operatorname{\text{\AA}}%
$, the empirical limits on the post-Newtonian potential determined by the
spectroscopic data are stronger than those extracted from the MTV-G scattering experiment.

At this point, we would like to mention that in a recent measurement of
$2S-4P$ transition frequency, the splitting between the states $4P_{3/2}$ and
$4P_{1/2}$ was determined with a precision of $4.3$ kHz \cite{2s4p}. In
principle, this more accurate measurement could be used to set a better
constraint for the post-Newtonian potential.

Another perspective to test modifications of gravity in the atomic length
scale is to consider the spectroscopy of heavier atoms or ions, especially
that of the element He whose transitions are being measured with a relative
precision of the order of $10^{-12}$ \cite{He}, and, for this reason, are
being used to probe new physics in the microscopic domain \cite{He1}.

Finally, we would like to highlight that a supposed modified gravitational
interaction would affect the isotope shift of atomic transitions. Therefore,
taking advantage of the well-studied theoretical framework of isotope shift
spectroscopy as well as the precision measurements of the correspondent
frequencies \cite{iso}, we could, from this complementary method, set new
constraints for the strength of non-standard gravitational interactions in the
atomic domain \cite{iso,iso1}.

\section{Appendix}

Here we compare the magnitude order of the indirect contribution of the
gravitational field through the Hamiltonian $H_{Eso}$(\ref{HEso}) and the
direct contribution given by $H_{Gso}$ to the fine structure of $P$ -states,
as discussed in section III.

Substituting (\ref{phiE}) in (\ref{HEso}), we may verify that the
gravitational correction of $H_{Eso}$ is given by:%
\begin{equation}
H_{Eso}^{(G)}=\frac{q^{2}}{8\pi\varepsilon_{0}m^{2}c^{2}r^{3}}(\varphi
/c^{2}+\tilde{\varphi}/c^{2})\left(  \vec{S}\cdot\vec{L}\right)  .
\end{equation}
Now let us write the potential $\varphi$ explicitly in terms of $r$ in both
Hamiltonians $H_{Gso}$ and $H_{Eso}^{(G)}$ . Considering just the Yukawa-term,
since the standard part is negligible, we obtain the following expressions in
magnitude order:%
\begin{align*}
&  H_{Eso}^{(G)}\symbol{126}\frac{q^{2}}{8\pi\varepsilon_{0}m^{2}c^{4}}%
\frac{\alpha GM}{r^{4}}e^{-r/\lambda}\left(  \vec{S}\cdot\vec{L}\right)  ,\\
H_{Gso}  &  \sim\frac{1}{mc^{2}}\left(  1+\frac{r}{\lambda}\right)
\frac{\alpha GM}{r^{3}}e^{-r/\lambda}(\vec{S}\cdot\vec{L}).
\end{align*}
For the sake of simplicity, here we have assumed $\alpha\sim\tilde{\alpha}$.
Now taking the average of the Hamiltonians in the 2P-state, for instance, we
obtain the following relation:%
\[
\left\langle H_{Eso}^{(G)}\right\rangle \sim\frac{(q^{2}/4\pi\varepsilon
_{0}a_{0})}{mc^{2}}\left(  \frac{\lambda}{\lambda+a_{0}}\right)  \left\langle
H_{Gso}\right\rangle .
\]
Notice that the coefficient is proportional to the ratio between the energy of
the hydrogen ground state and the rest energy of the electron, which is of the
order of $10^{-5}$. It also depends on a certain relation between $\lambda$
and $a_{0}$, which is lesser than $1$ for all value of $\lambda$.

\begin{acknowledgement}
G. C. Luna thanks CAPES for financial support.
\end{acknowledgement}

\end{document}